\documentclass{aip-cp}

\usepackage[numbers]{natbib}
\usepackage{rotating}
\usepackage{graphicx}
\usepackage{mathrsfs}
\usepackage{amssymb}
\usepackage{epsf}
\usepackage{epstopdf}
\usepackage[normalem]{ulem}
\usepackage{color, soul}
\usepackage[dvipsnames]{xcolor}
\usepackage{bm}
\usepackage{times}

\renewcommand\sout{\bgroup \color{red} \ULdepth=-.5ex \ULset}

\renewcommand{\rm}[1]{\textrm{#1}}

\begin{document}

\title{Quark matter symmetry energy effect on equation of state for neutron stars}
\author[aff1,aff2]{Xuhao Wu}
\author[aff2]{Akira Ohnishi}
\corresp{E-mail: wuhaobird@gmail.com, ohnishi@yukawa.kyoto-u.ac.jp, shennankai@gmail.com}
\author[aff1]{Hong Shen}
\affil[aff1]{School of Physics, Nankai University, Tianjin 300071, China}
\affil[aff2]{Yukawa Institute for Theoretical Physics, Kyoto University, Kyoto 606-8502, Japan}
\maketitle

\begin{abstract}
We study the equation of state of neutron-star matter including the effects of isovector-vector coupling
in quark matter. We employ the relativistic mean-field theory with an extended TM1 parameter set to describe hadronic matter and the SU(3) Nambu-Jona-Lasinio model with isovector-vector and hypercharge-vector couplings to describe the quark matter. The deconfinement hadron-quark phase transition is constructed under Gibbs phase equilibrium conditions.
It is found that the isovector-vector and hypercharge-vector couplings in quark matter enhance the symmetry energy and hypercharge symmetry energy in neutron-star matter.
\end{abstract}

\section{INTRODUCTION}\label{sec1}
The study of equation of state (EOS) has attracted much attention because it is a crucial input
for modeling neutron stars and supernovae~\cite{Latt85,Latt91,Shen11}. In the inner core of
neutron stars, the baryon number density $n_b$ may reach 5-10 times nuclear saturation density $n_0$,
where the deconfinement hadron-quark phase transition may occur~\cite{Glen01,Heis00,Webe05}.
Because there does not exist a unified model that can describe both hadronic matter and deconfined quark matter, a realistic method is to consider these two phases in different models. The relativistic mean-field (RMF) theory is a commonly used method to investigate nuclear matter and finite nuclei. In the RMF model,
the isovector-vector meson ($\rho$) plays an essential role to control the symmetry energy $S_0$ and its slope $L$ in hadronic matter, which are important for understanding many phenomena in nuclear physics and astrophysics~\cite{Liba08,Horo01,Oyama07,Ducoin10}. The symmetry energy at saturation density is constrained by experiments to be about $30  \leq S_0 \leq 32$ MeV, and the slope parameter $L$ is limited in the range of $40 - 60$ MeV~\cite{Tews17}. In this work, we use an extended TM1 parameter set with $L=50$ MeV to describe hadronic matter~\cite{Bao14b}, where the model parameters were determined
by the properties of finite nuclei. On the other hand, the u-d quark isospin asymmetry plays an important role in the QCD phase diagram~\cite{Ohnishi2012,Dexheimer2018,Ueda2013} and quark matter symmetry energy\cite{Chu2013,Chu2017,Chu2015,Liu2016,Pagliara2010,Shao2012,Toro2006}.
We use the three flavor Nambu-Jona-Lasinio (NJL) model with isovector-vector
and hypercharge-vector couplings to describe the quark matter.
The isovector-vector coupling was chosen to be $G_3=1.5\ G_0$ in Refs~\cite{Wu17,Rehb96}, $G_3=G_0$ in Ref~\cite{Ueda13}, and $G_3=0$ in Ref~\cite{Benic:2014jia}, where $G_0$ is the isoscalar-vector coupling.
This difference comes from which kind of vector coupling terms is used, $(\bar{q}\gamma_\mu q)^2$
or  $\sum_\alpha \left[(\bar{q}\gamma_\mu \lambda_\alpha q)^2
+(\bar{q}i\gamma_\mu\gamma_5 \lambda_\alpha q)^2\right]$~\cite{Chu16}. The second one contains
the isovector-vector coupling, which can significantly affect the symmetry energy of quark matter.
In Ref.~\cite{Pereira16}, the authors used a small isovector-vector coupling in the NJL model, the resulting EOS was found to be sensitive to the isoscalar-vector coupling.
The neutron-star observations of PSR J1614-2230~\cite{Demo10,Fons16} and PSR J0348+0432~\cite{Anto13}
constrain that the neutron-star maximum mass should be larger than $2\ M_\odot$ , with $M_\odot$ being the solar mass. In Ref~\cite{Annala2017},
the authors showed that the radius of a 1.4 $M_\odot$ neutron star is in the range of $10\,\mathrm{km} \lesssim R_{1.4} \lesssim 13.6\,\mathrm{km}$.

\section{HADRON-QUARK PHASE TRANSITION IN NEUTRON STARS}\label{sec2}
\subsection{Hadronic Phase}
We employ the RMF theory to describe the hadronic phase, which contains
nucleons ($n$ and $p$) and leptons ($e$ and $\mu$).  In the RMF approach,
nucleons interact through the exchange of isoscalar-scalar ($\sigma$), isoscalar-vector ($\omega$),
and isovector-vector ($\rho$) mesons. The Lagrangian density is given by
\begin{eqnarray}
\label{eq:LRMF}
\mathcal{L}_{\rm{RMF}} & = & \sum_{i=p,n}\bar{\psi}_i
\left\{i\gamma_{\mu}\partial^{\mu}-\left(M+g_{\sigma}\sigma\right)
\right.
\left.-\gamma_{\mu} \left[g_{\omega}\omega^{\mu} +\frac{g_{\rho}}{2}\tau_a\rho^{a\mu}
\right]\right\}\psi_i  \nonumber\\
&& +\frac{1}{2}\partial_{\mu}\sigma\partial^{\mu}\sigma -\frac{1}{2}%
m^2_{\sigma}\sigma^2-\frac{1}{3}g_{2}\sigma^{3} -\frac{1}{4}g_{3}\sigma^{4}
\nonumber\\
&& -\frac{1}{4}W_{\mu\nu}W^{\mu\nu} +\frac{1}{2}m^2_{\omega}\omega_{\mu}%
\omega^{\mu} +\frac{1}{4}c_{3}\left(\omega_{\mu}\omega^{\mu}\right)^2  \nonumber
\\
&& -\frac{1}{4}R^a_{\mu\nu}R^{a\mu\nu} +\frac{1}{2}m^2_{\rho}\rho^a_{\mu}%
\rho^{a\mu}
\nonumber\\&&
+\Lambda_{\rm{v}} \left(g_{\omega}^2
\omega_{\mu}\omega^{\mu}\right)
\left(g_{\rho}^2\rho^a_{\mu}\rho^{a\mu}\right) \nonumber\\
&& +\sum_{l=e,\mu}\bar{\psi}_{l}
  \left( i\gamma_{\mu }\partial^{\mu }-m_{l}\right)\psi_l,
\end{eqnarray}
where $W^{\mu\nu}$ and $R^{a\mu\nu}$ are the antisymmetric field tensors 
for $\omega^{\mu}$ and $\rho^{a\mu}$, respectively. Under the mean-field approximation,
the non-vanishing expectation values of meson fields are denoted as
$\sigma=<\sigma>$, $\omega=<\omega^0>$, and  $\rho=<\rho^{30}>$.
From the Lagrangian density, we derive the equations of motion for nucleons and mesons.
These coupled equations, together with the $\beta$ equilibrium condition,
can be solved self-consistently, and the properties of hadronic matter are obtained
at a given baryon density. In this work, we use an extended TM1 parameter set~\cite{Bao14b}, named as TM1-50, which was generated by adjusting $g_{\rho}$ and $\Lambda_v$ simultaneously so as to
achieve $L=50$ MeV at the saturation density $n_0$ and keeping the symmetry energy
$S_0$ fixed at a density of $0.11$ fm$^{-3}$. The TM1-50 parameter set is given in Table~\ref{tab:para}.
We do not include hyperon degree of freedom due to the hyperon puzzle. Considering repulsive interaction at high densities or using crossover mixed phase with earlier onset may help avoid this problem, then the hyperon effect should be small because of low hyperon fractions.

\begin{table*}[hpbt]
\caption{TM1-50 parameter set.
The masses are given in the unit of MeV.}
\begin{center}
\footnotesize
\begin{tabular}{lcccccccccccc}
\hline
Model   &$L$(MeV) &$M$  &$m_{\sigma}$  &$m_\omega$  &$m_\rho$  &$g_\sigma$  &$g_\omega$
        &$g_\rho$ &$g_{2}$ (fm$^{-1}$) &$g_{3}$ &$c_{3}$  &$\Lambda_{\rm{v}}$\\
\hline
TM1-50    &50   &938.0  &511.198  &783.0  &770.0  &10.0289  &12.6139  &12.2413
        &$-$7.2325   &0.6183   &71.3075   &0.0327\\
\hline
\end{tabular}
\end{center}
\label{tab:para}
\end{table*}

\subsection{Quark Phase}
The three flavor NJL model is adopted for the quark phase.
The Lagrangian density is given by
\begin{eqnarray}
\label{eq:Lnjl}
\mathcal{L}_{\rm{NJL}} &=&\bar{q}\left( i\gamma _{\mu }\partial ^{\mu
}-m^{0}\right) q+{G_S}\sum\limits_{a = 0}^8 {\left[ {{{\left( {\bar q{\lambda _a}q} \right)}^2}
+ {{\left( {\bar q i{\gamma _5}{\lambda _a}q} \right)}^2}} \right]}  \nonumber \\
&&-K\left\{ \det \left[ \bar{q}\left( 1+\gamma _{5}\right) q\right] +\det %
\left[ \bar{q}\left( 1-\gamma _{5}\right) q\right] \right\} +\mathcal{L}_{V}
 ,\nonumber \\
\end{eqnarray}%
with
\begin{eqnarray}
{{\cal L}_V}
=& - {G_0}{\left( {\bar q{\gamma ^\mu }q} \right)^2}
- {G_V}\sum\limits_{\alpha=1}^8
      \left[ \left(\bar{q} \gamma ^\mu        \lambda_\alpha q\right)^2
           + \left(\bar{q}i\gamma ^\mu\gamma_5\lambda_\alpha q\right)^2
	\right]
\ ,
\end{eqnarray}
in which $q$ denotes the quark field with three flavors ($N_f$=3) and three colors ($N_c$=3).
$G_S$, $G_0$, and $G_V$ are the scalar, flavor-singlet-vector, and flavor-octet-vector coupling constants, respectively.
Under the mean-field approximation, only the terms with diagonal matrix elements in $\lambda_\alpha$ remain, and as a result, ${\cal L}_V$ is reduced to
\begin{eqnarray}
{{\cal L}_V} &=&  - {G_0}{\left( {\bar q{\gamma ^\mu }q} \right)^2} - {G_3}\left[ {{{\left( {\bar q{\gamma ^\mu }{\lambda _3}q} \right)}^2} + {{\left( {\bar qi{\gamma ^\mu }{\gamma _5}{\lambda _3}q} \right)}^2}} \right] \nonumber \\
&&- {G_8}\left[ {{{\left( {\bar q{\gamma ^\mu }{\lambda _8}q} \right)}^2} + {{\left( {\bar qi{\gamma ^\mu }{\gamma _5}{\lambda _8}q} \right)}^2}} \right].
\end{eqnarray}
We consider the quark matter in $\beta$ equilibrium, which includes quarks ($u$, $d$ and $s$) and leptons ($e$ and $\mu$).
We employ the parameter set given in Ref.~\cite{Rehb96}, $m_{u}^{0}=m_{d}^{0}=5.5\ $MeV,
$m_{s}^{0}=140.7\ $MeV, $\Lambda =602.3\ $MeV, ${G_S}\Lambda^{2}=1.835$,
and $K\Lambda ^{5}=12.36$. The vector couplings ($G_0, G_3, G_8$) are treated as free
parameters. It is found that a large $G_0$ value ($G_0 > 0.27\,G_S$) leads to a high energy density of quark matter, and as a result, the deconfinement phase transition will not occur. Therefore, we use $G_0=0.25\,G_S$ and $G_3=G_8=(0,\,1.5,\,10)\ G_0$.
For the case of $G_3=G_8=1.5\ G_0$, it corresponds to the Lagrangian adopted in Refs.~\cite{Wu17,Rehb96},
which is the flavor SU(3) limit. The parameter choice $G_3=G_8=10\ G_0$ gives a
symmetry energy slope of $L_Q\simeq 50$ MeV in quark phase, which is comparable to the slope obtained by the TM1-50 parameter set in hadronic phase.
The vector couplings increase the energy per baryon as
\begin{eqnarray}
\label{eq:enev}
\frac{\Delta\varepsilon_V}{n_b}
=&9\,G_0\,n_b
+G_3\,n_b\,\delta^2
+3\,G_8\,n_b\,\delta_h^2
\ ,\\
\delta=& \frac{n_d-n_u}{n_b}
\ ,\quad
\delta_h= \frac{n_b-n_s}{n_b}=\frac{B+S}{B}=\frac{Y}{B}
\ .
\end{eqnarray}
$\delta$ and $\delta_h$ indicate the isospin asymmetry and the hypercharge ($Y$) fraction.
Since the symmetry energy is defined as the coefficient of $\delta^2$, the symmetry energy contributed
from the vector couplings is given by
\begin{eqnarray}
\Delta S_V(n_b)=&G_3\,n_b
=G_3\,n_0 + 3\,G_3\,n_0\,\left(\frac{n_b-n_0}{3n_0}\right)
\ .
\end{eqnarray}
The vector coupling contribution to the slope parameter is $\Delta L_V=3\,G_3\,n_0=6.6$ and 44 MeV
for $G_3=1.5\ G_0$ and $10\ G_0$, respectively.
The parameter choice $G_3=G_8=(0-10)\,G_0$ covers a wide range of the symmetry energy.

\subsection{Hadron-Quark Phase Transition}
We adopt the Gibbs construction for the mixed phase which connects the pure hadronic phase and pure quark phase.
For the mixed phase presented in neutron stars, the two coexisting phases satisfy the conditions of global charge neutrality, $\beta$ equilibrium, and mechanical equilibrium, which are written as
\begin{eqnarray}
&&un_c^{{\rm{QP}}} + \left( {1 - u} \right)n_c^{{\rm{HP}}} = 0
,  \\
&&\mu_u+\mu_e=\mu_d=\mu_s=\frac{1}{3}\left(\mu_n+\mu_e\right), \\
&&{P_{{\rm{HP}}}}\left( {{\mu _n},{\mu _e}} \right) = {P_{{\rm{QP}}}}\left( {{\mu _n},{\mu _e}} \right),
\end{eqnarray}%
where $u=V_{\rm{QP}}/(V_{\rm{QP}}+V_{\rm{HP}})$
represents the volume fraction of quark matter in the mixed phase. $n_c^{{\rm{QP}}}$ and $n_c^{{\rm{HP}}}$
represent the charge number density of quark phase and hadronic phase, respectively.
By using these equilibrium conditions, we calculate the properties of the
hadron-quark mixed phase at a given baryon density
$n_b=un_b^{{\rm{QP}}} + \left( {1 - u} \right)n_b^{{\rm{HP}}}$.

\section{RESULTS AND DISCUSSION}
In this section, we study the effects of quark-matter symmetry energy on the EOS and neutron-star properties. To achieve this purpose, we use the RMF model to describe hadronic matter and the NJL with isovector-vector and hypercharge-vector couplings to describe the quark matter.
The Gibbs construction is employed for the hadron-quark mixed phase. In the NJL model,
the isovector-vector couplings are related to the quark-matter symmetry energy, which
can affect the hadron-quark phase transition.
In the present work, we use the isoscalar-vector coupling $G_0=0.25\ G_S$, and compare the results for
$G_3=G_8=(0, 1.5, 10)\ G_0$.
\begin{figure*}[tbhp]
\centering
\includegraphics[bb=0 0 580 420, width=5 cm,clip]{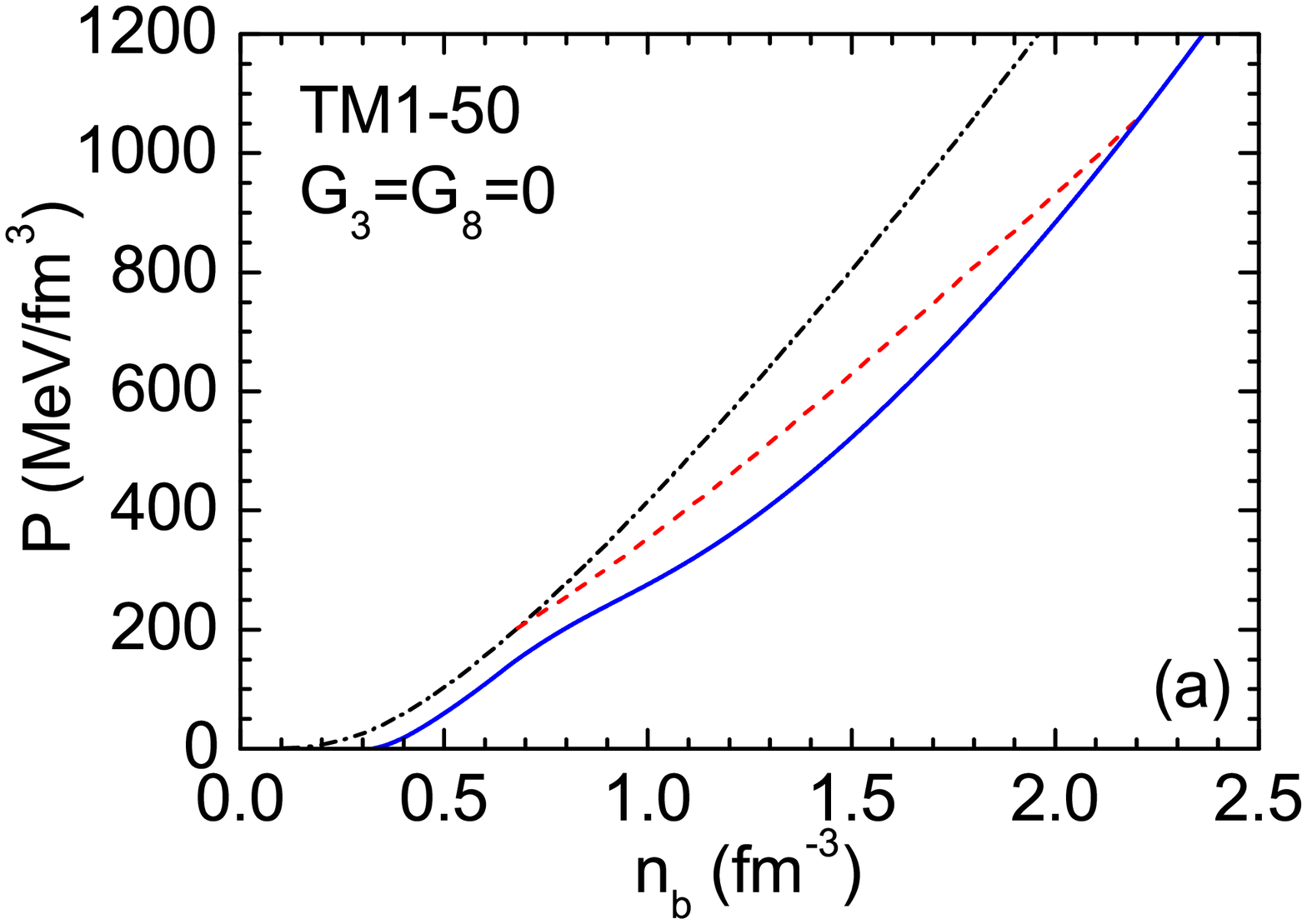}
\includegraphics[bb=0 0 580 420, width=5 cm,clip]{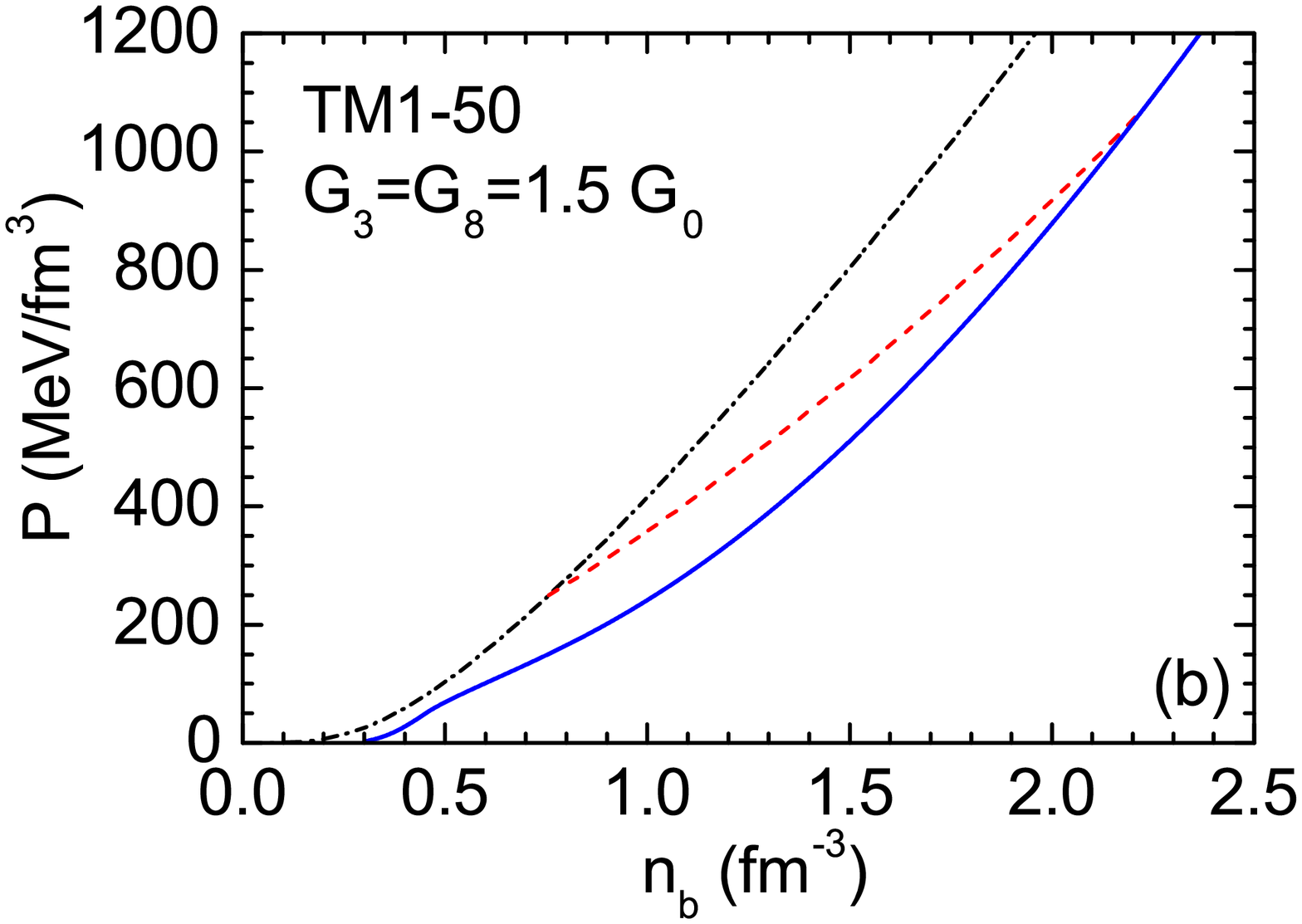}
\includegraphics[bb=5 5 580 420, width=5 cm,clip]{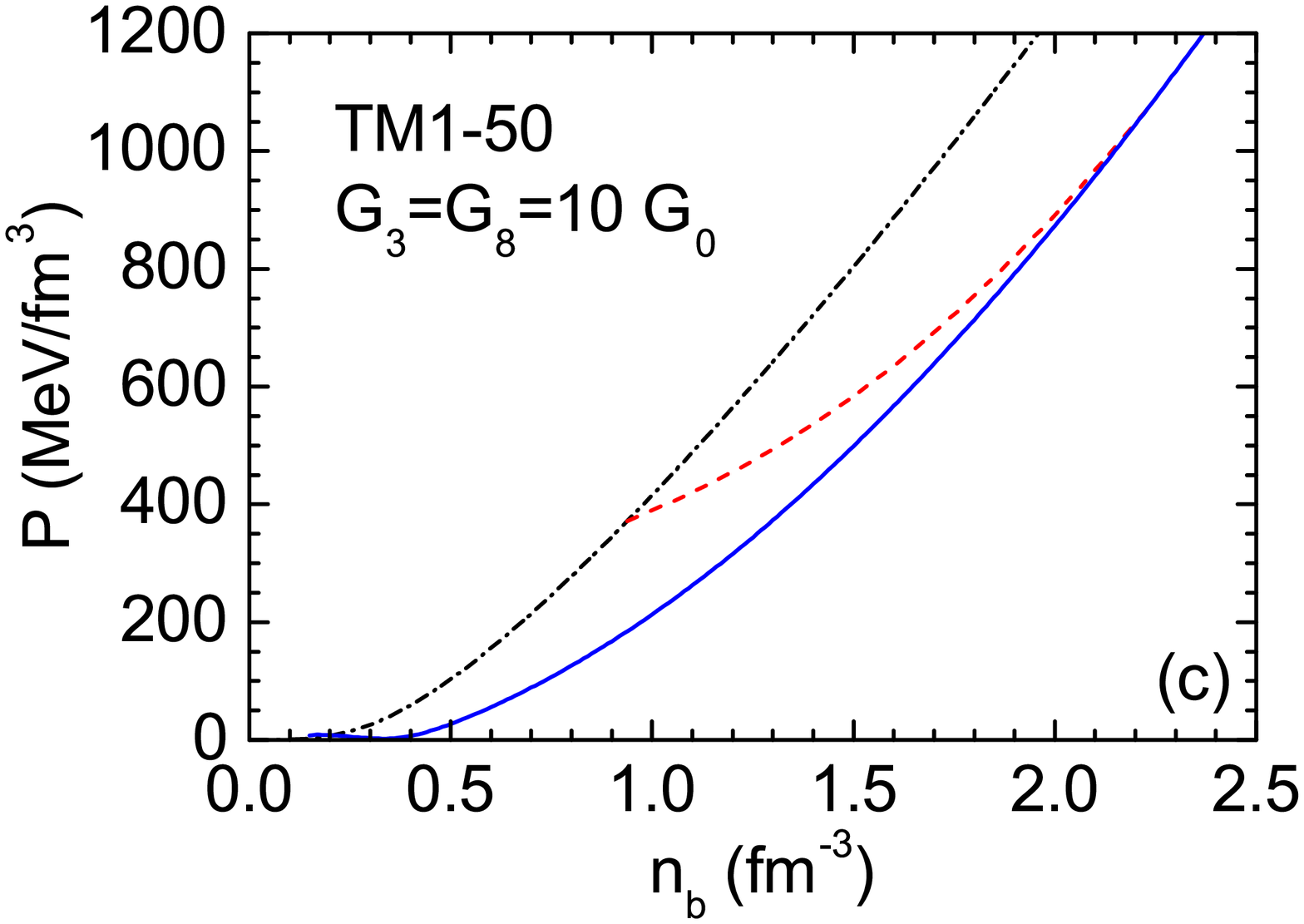}
\caption{(Color online) Pressures $P$ as a function of the baryon number density $n_b$
for the pure hadronic phase (dash-dotted lines), the mixed phase (dashed lines)
and the pure quark phase (solid lines).}
\label{fig:1nbp}
\end{figure*}
\begin{figure*}[hptb]
\centering
\includegraphics[bb=0 0 580 580, width=7 cm,clip]{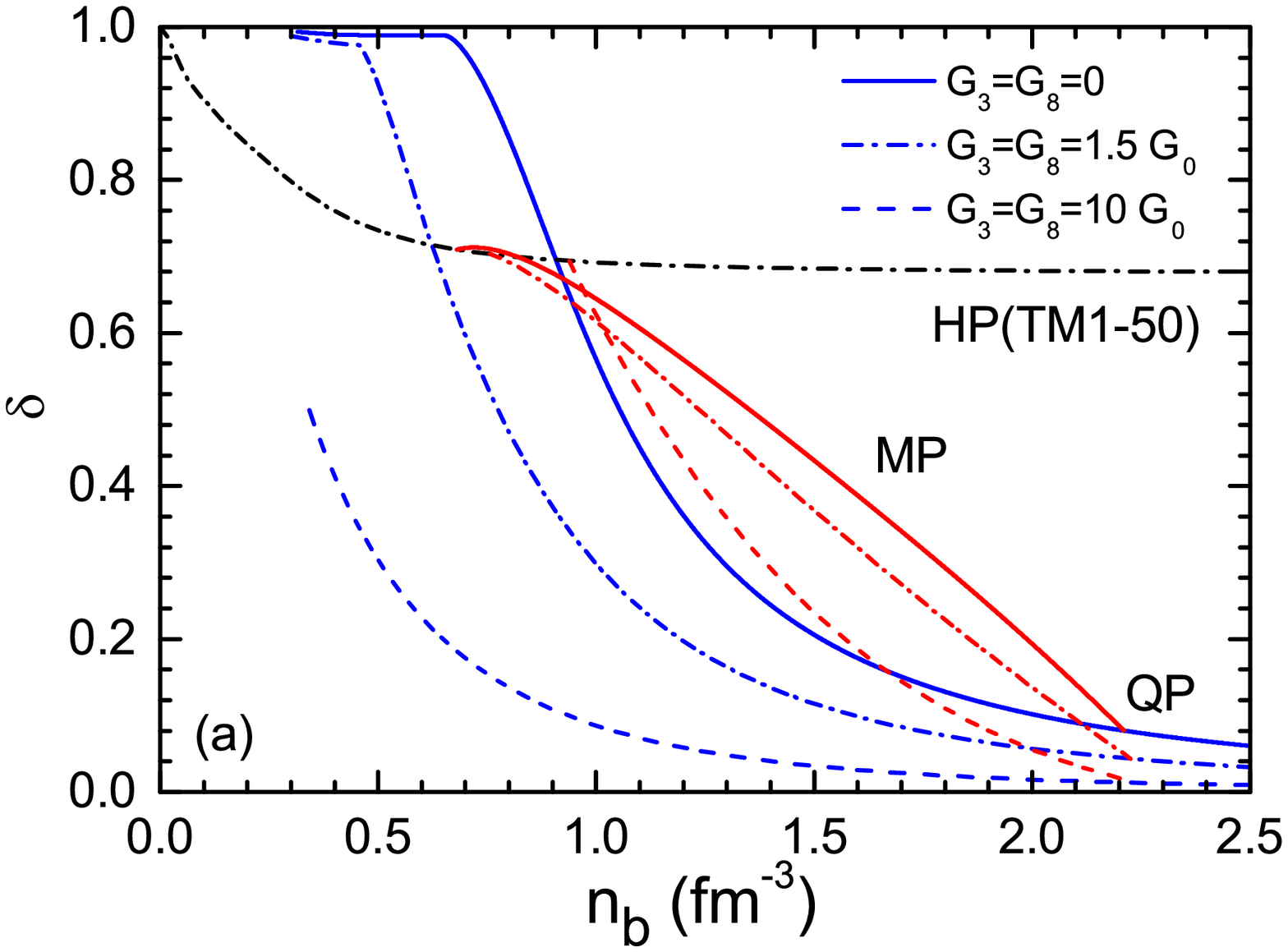}
\includegraphics[bb=0 0 580 580, width=7 cm,clip]{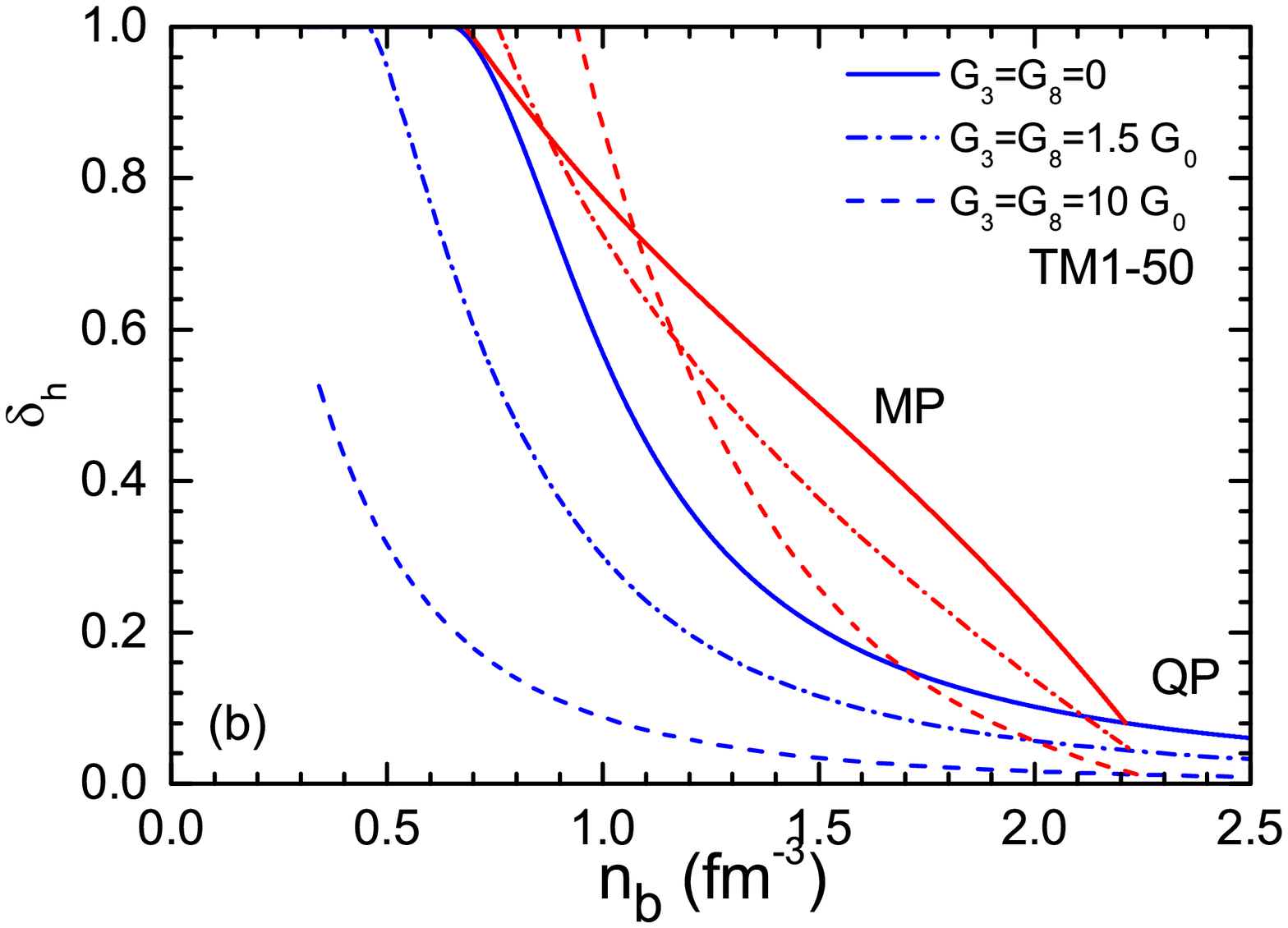}
\caption{(Color online) The isospin asymmetry $\delta$ and hypercharge fraction $\delta_h$ as a function of
the baryon number density $n_b$.}
\label{fig:2delta}
\end{figure*}
\begin{table*}[bthp]
\caption{Parameter dependence of the phase transition densities
and neutron star properties, $G_0\,=\,0.25\,G_S$. Number densities are
given in the unit of fm$^{-3}$ and radius are given in the unit of km.}
\begin{center}
\begin{tabular}{lcccccccccccc}
\hline\hline
Model &$G_3/G_0$ &$G_8/G_0$  &$n^{(1)}_{b}$  &$n^{(2)}_{b}$  &${M_{\rm{max}}}/{M_\odot}$  &$n_c^{M_{\rm{max}}}$
        &$R(1.4\ M_\odot)$ & $R({M_{\rm{max}}})$\\
\hline
TM1-50 &-  &-      &-        &-                      & 2.120 & 0.899  &13.0  & 11.73 \\
&0    &0    &0.681  &2.210             	       &2.101  &0.900  &13.0 & 12.01\\
TM1-50/NJL&1.5  &1.5  &0.757  &2.226  &2.112   &0.879  &13.0 & 12.01\\
&10   &10   &0.938  &2.241                      &2.120   &0.888  &13.0 & 11.73\\
\hline
&  &  &  &  & 1.928 $\pm$ 0.017 ~\cite{Demo10,Fons16}    &  &12 $\pm$ 1 ~\cite{Steiner16}& \\
Constraints &-  &-      &-  &-      & 2.01 $\pm$ 0.04 ~\cite{Anto13} &-  &9.4 $\pm$ 1.2 ~\cite{Guillot14} & -\\
&  &  &   &   &  &  &$>14$~\cite{Haensel:2016th} &\\
\hline\hline
\end{tabular}
\label{tab:2coex}
\end{center}
\end{table*}

To check the effect of the vector couplings on the EOS of neutron-star matter, we show the pressure $P$
as a function of the baryon number density $n_b$ in Fig.\ref{fig:1nbp}.
The left, middle, and right panels show the results of $G_3=G_8=0, 1.5, 10\ G_0$, respectively.
It is found that the density range of the mixed phase is clearly dependent on the parameter used.
We define $n^{(1)}_b$ and $n^{(2)}_b$ as the starting and the ending densities of the mixed phase, respectively.
The results with different couplings are listed in Table \ref{tab:2coex}. The visible delay of $n_{(1)}^b$ comes from the energy increase of the quark phase given by Eq.~(\ref{eq:enev}), however this effect becomes smaller with increasing density $n_b$. This is because that the isospin asymmetry $\delta=(n_d-n_u)/(n_b)$ and the hypercharge fraction $\delta_h=(n_b-n_s)/(n_b)$ depend on the baryon density $n_b$,
as shown in Fig.\ref{fig:2delta}. The quark densities, $n_u$, $n_d$, and $n_s$, become close with each other as $n_b$ increases. This trend is relatively weak in the mixed phase due to the negative charge of quark matter.

\begin{figure*}[hptb]
\centering
\includegraphics[bb=0 0 580 580, width=7 cm,clip]{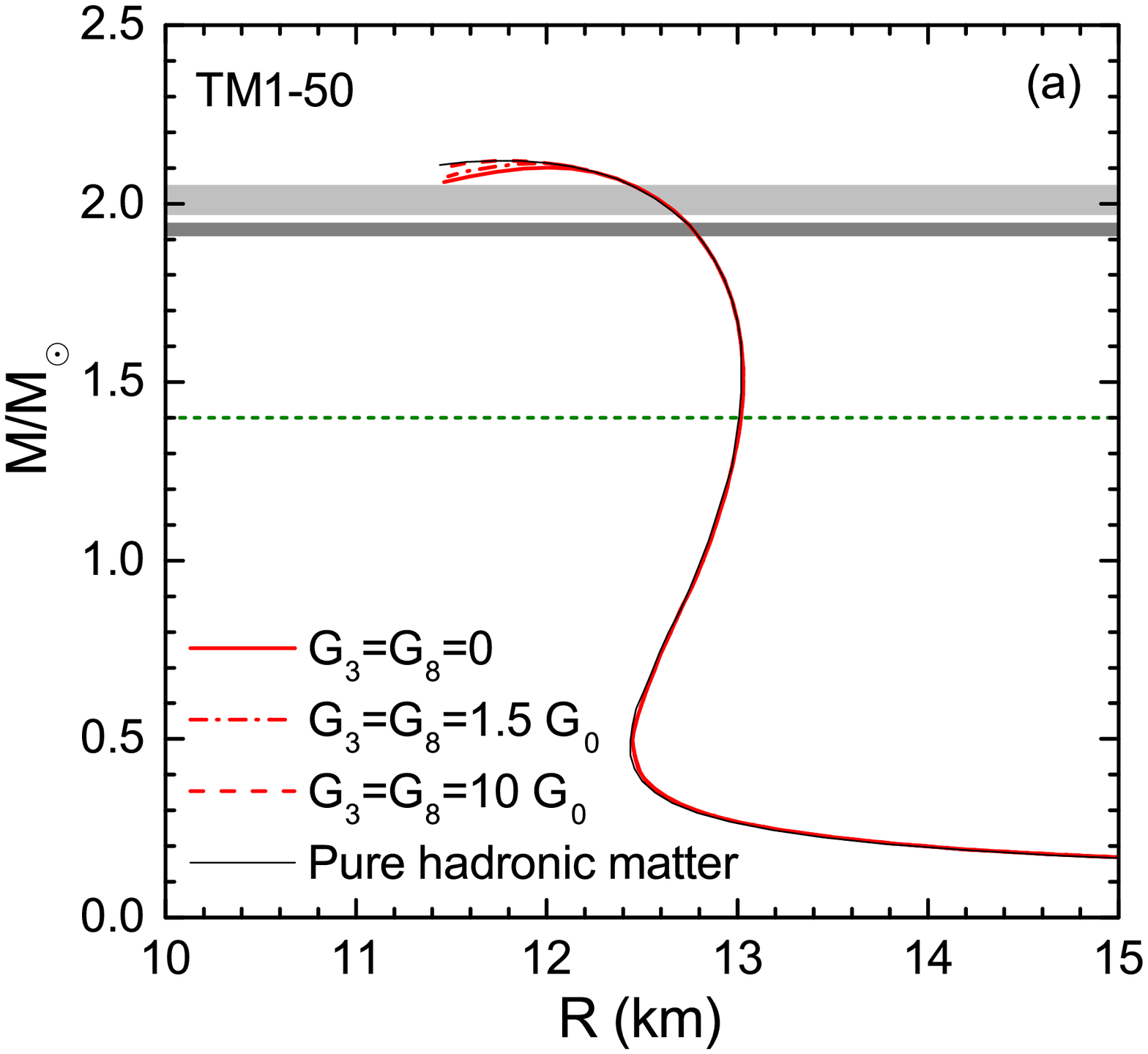}
\includegraphics[bb=0 0 580 580, width=7 cm,clip]{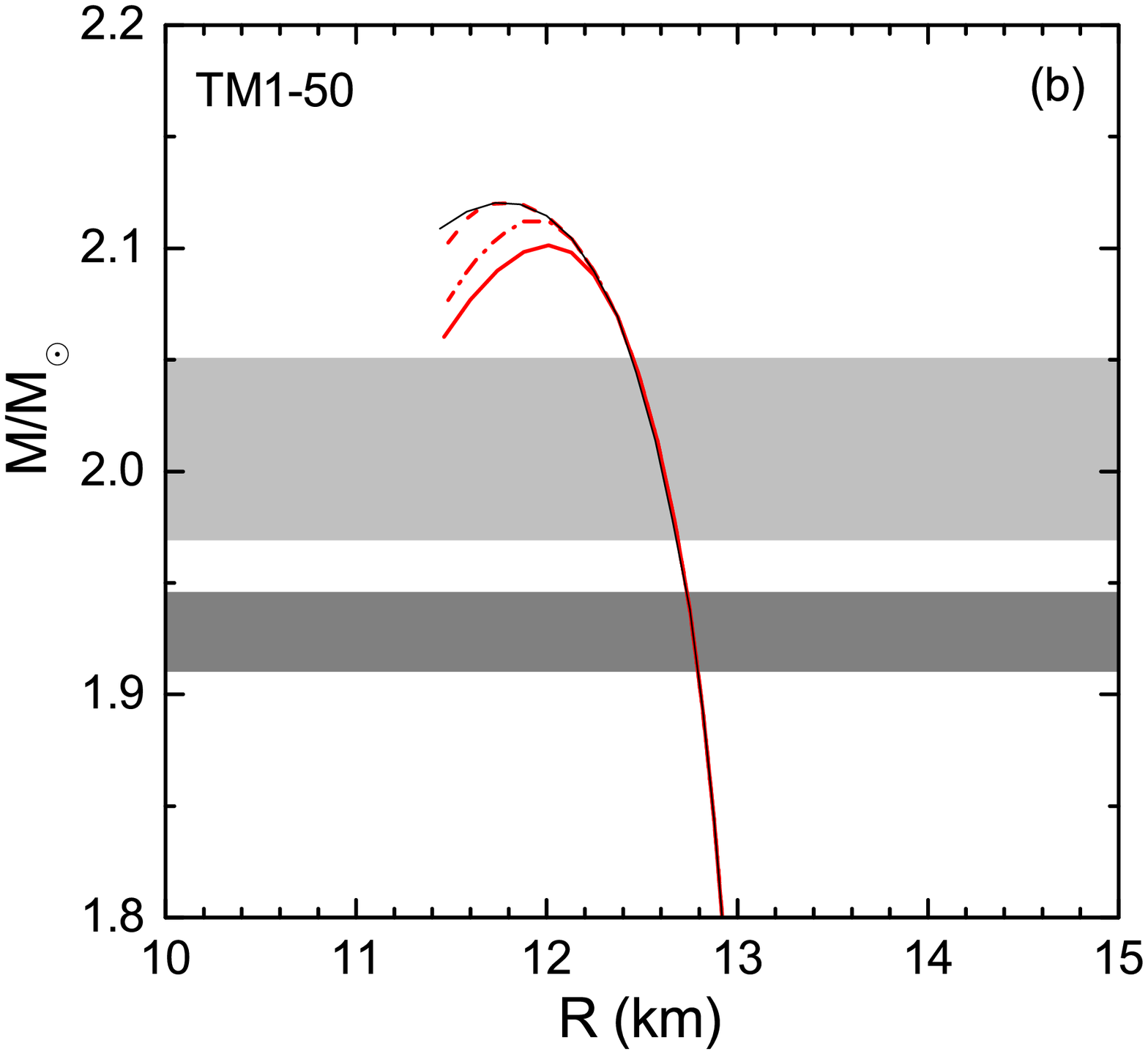}
\caption{(Color online) The left panel plots mass-radius relations
of neutron stars for different EOS and the right panel is an enlargement.}
\label{fig:3rm}
\end{figure*}

The neutron-star observations of PSR J1614-2230 and PSR J0348+0432
provide a constraint on the maximum mass of neutron stars, namely, the maximum mass should be larger than
$2\ M_\odot$. By solving the Tolman-Oppenheimer-Volkoff (TOV) equation with the EOS obtained, we can examine the effect of the isovector-vector coupling $G_3$ and hypercharge-vector coupling $G_8$ on the properties of neutron stars. In Fig.\ref{fig:3rm},
we show the neutron star mass-radius
relations with different coupling constants. The thin black solid line shows the result of pure hadronic
matter, which predicts a maximum neutron-star mass of $2.12\ M_{\odot}$. When the quark degree of freedom is included, the EOS becomes soft and the maximum mass slightly decreases. The inclusion of the isovector-vector and hypercharge-vector couplings in the quark phase can attenuate this effect, as shown
in Fig.\ref{fig:3rm}. We list the central density and maximum mass of neutron stars obtained with different conditions in Table~\ref{tab:2coex}. It is found that the central density $n_c$ is less than the transition density $n^{(1)}_b$ for
$G_3=G_8=10\ G_0$, which means the hadron-quark phase transition does not occur in neutron stars for such high isovector-vector and hypercharge-vector couplings.

In summary, the effects of isovector-vector and hypercharge-vector couplings in the NJL model on the
hadron-quark phase transition and neutron-star properties were investigated. We employed the RMF theory to describe hadronic matter and the NJL model to describe the quark matter.
We used the Gibbs construction for the mixed phase that connects the pure hadronic phase and pure quark phase. 
We found that by including isovector-vector and hypercharge-vector couplings in the NJL model, the hadron-quark
phase transition delayed and shrank. By using the EOS obtained, the resulting properties of neutron stars are influenced by the isovector-vector and hypercharge-vector couplings, which slightly increase the maximum mass of neutron stars. With small isovector-vector and hypercharge-vector couplings, a relatively large mixed core is achieved.
When the coupling constant is as large as $G_3=G_8=10\ G_0$, only pure hadronic matter exists in neutron stars. \\

\section{ACKNOWLEDGMENTS}
This work was supported in part by
the National Natural Science Foundation of China (No. 11675083),
the Grants-in-Aid for Scientific Research from JSPS
(Nos. 15K05079, 15H03663, 16K05350),
the Grants-in-Aid for Scientific Research on Innovative Areas from MEXT
(Nos. 24105001, 24105008),
and by the Yukawa International Program for Quark-hadron Sciences (YIPQS).


\end{document}